\documentclass[12pt]{article}
\usepackage[dvipdfmx]{hyperref}
\usepackage{epsfig}
\usepackage{url}
\makeatletter
\def\alt{\mathrel{\mathpalette\gl@align<}}
\def\agt{\mathrel{\mathpalette\gl@align>}}
\def\gl@align#1#2{\lower.6ex\vbox{\baselineskip\z@skip\lineskip\z@
\ialign{$\m@th#1\hfil##\hfil$\crcr#2\crcr\sim\crcr}}}
\makeatother

\newcommand{\doi}[1]{\href{https://doi.org/#1}{doi:#1}}

\begin{document}
\begin{flushright}
MI-TH-1810\\
November, 2018
\end{flushright}
\vspace*{1.0cm}
\begin{center}
\baselineskip 20pt
{\Large\bf
Electron $g-2$ with flavor violation in MSSM} \vspace{1cm}

{\large
Bhaskar Dutta$^1$ and Yukihiro Mimura$^2$}
\vspace{.5cm}

$^1${\it
Department of Physics, Texas A\&M University,
College Station, TX 77843-4242, USA
}\\
$^2${\it
Institute of Science and Engineering, Shimane University, Matsue 690-8504, Japan
}
\\

\vspace{.5cm}

\end{center}

\begin{center}{\bf Abstract}\end{center}

The muon $g-2$ anomaly is a long-standing discrepancy from its
standard model prediction. The recent improved measurement of the fine structure 
constant makes the electron $g-2$ anomaly  very interesting for both sign and magnitude in comparison
to the muon $g-2$ anomaly.
In order to explain both muon and electron $g-2$ anomalies, we
introduce  flavor violation in the minimal supersymmetric standard model (MSSM)
as a low energy theory.
The lepton flavor violating process $\tau \to e\gamma$
is one the major constraints to explain both $g-2$ anomalies simultaneously emerging from flavor violating terms.
We show various  mass spectra of  sleptons, neutralinos, and charginos,
consistent with the LHC results,
to explain both anomalies after satisfying the  flavor violation constraint.

\thispagestyle{empty}

\bigskip
\newpage

\addtocounter{page}{-1}

\section{Introduction}
\baselineskip 18pt

The anomalous magnetic moment ($g-2$) of muon is one of the long-standing 
deviations from its standard model (SM) prediction.
The discrepancy between the experiment  \cite{Bennett:2006fi,Tanabashi:2018oca}
 and the SM prediction \cite{Blum:2018mom,Keshavarzi:2018mgv} of
$a_\mu = (g-2)_\mu /2$ is more than $3.5\sigma$:
\begin{equation}
\Delta a_\mu = (2.74 \pm 0.73) \times 10^{-9}.
\end{equation}
The SM prediction is smaller than the experimental measurement. There will be new measurements of 
$a_\mu$ at Fermilab very soon and at J-PARC. The theory prediction is expected to have a better accuracy.

On the other hand, 
the electron $g-2$
has been consistent \cite{Aoyama:2017uqe,Hanneke:2008tm} with the measurement.
However, a 
recent  precise measurement of the fine structure constant 
\cite{Parker}
has changed the situation,
which leads to a  $2.4\sigma$ discrepancy  in the electron $g-2$ \cite{Davoudiasl:2018fbb}
\begin{equation}
\Delta a_e = (-8.7 \pm 3.6) \times 10^{-13}.
\end{equation}
The SM prediction is larger  than the experimental measurement in this case,
and the sign is opposite.
Without any flavor violation in the lepton sector,
the anomalous magnetic moments of electron and muon obey the 
lepton mass scaling as
\begin{equation}
\Delta a_e/\Delta a_\mu = m_e^2/m_\mu^2 \simeq 2.4 \times 10^{-5}, \label{LMS}
\end{equation}
even if there is an effect from the physics beyond SM.
In that sense, both sign and magnitude have discrepancies.
The theoretical implication has been studied on this issue
\cite{Davoudiasl:2018fbb,Crivellin:2018qmi,Liu:2018xkx}.

The minimal supersymmetric (SUSY) standard model (MSSM)
is one of the promising candidates of the models beyond SM. However the SUSY particles have not yet been observed at the LHC. Based on the recent results, the colored SUSY particles, e.g., squarks ($\tilde q$), gluinos ($\tilde g$)  are heavier than 1.6 - 2 TeV~\cite{Sirunyan:2017kqq,Aaboud:2017vwy}.
 However the constraints on the  non-colored sparticles, e.g., sleptons  ($\tilde \ell$), charginos ($\tilde \chi^+$), 
 neutralinos ($\tilde \chi^0$) etc.  are not very good~\cite{cmsstop,atlasstop} and a lot parameter space is available in the mass range 100 GeV to 1 TeV for these particles. This mass range is very important for these particles to contribute to  the $g-2$ calculations. 
 
The muon $g-2$ anomaly has been studied in the context of 
MSSM \cite{Lopez:1993vi,Everett:2001tq,Crivellin:2011jt,Endo:2013bba,Gogoladze:2014cha,Ajaib:2015yma,Chowdhury:2015rja,Hagiwara:2017lse,Bhattacharyya:2018inr}.
In MSSM, there can be a loop diagram in which
a slepton and a chargino (neutralino) propagate,
and it can explain the muon $g-2$ anomaly
provided the sleptons and chargino (neutralino) are adequately light, say less than a TeV. The LHC constraints are discussed in~\cite{Ajaib:2015yma} and it was shown that a large region of parameter space is still available and a sizable  parameter space will still be available even  after the LHC  acquires 3000 fb$^{-1}$ of luminosity.

However, even if  the central value of muon $g-2$ can be explained by the sparticle loop diagrams,
the electron $g-2$ contribution only provides $\Delta a_e \simeq 6.5 \times 10^{-14}$
if there is no flavor violation.
In this paper, we consider the MSSM with flavor violation as a weak scale theory.
We show that the major constraint to reproduce the central value of the electron $g-2$
is $\tau \to e\gamma$,
and we study if  $g-2$ anomalies of muon and electron can both be accommodated
after satisfying the $\tau \to e\gamma$ constraint.
The mass spectrum to reproduce the electron $g-2$ depends on the choice of flavor violation,
and 
the slepton masses need to be adequately light. 
In this work we will use the masses of sleptons, charginos, and neutralinos 
allowed by the LHC results to explain the electron and muon $g-2$ anomalies.

This paper is organized as follows. In section 2, we discuss the possible explanations of the  electron $g-2$ anomaly in the context of the MSSM and associated flavor violations. In section 3, we describe our numerical fit of both electron and muon $g-2$ anomalies  satisfying  the LHC  and the flavor violation constraints and section 5 contains our conclusion.

\section{Electron $g-2$ and flavor violation}

In the MSSM, the chargino loop diagram 
with Higgsino-wino propagation ($\tilde H$-$\tilde W$) gives the dominant contribution
to the anomalous magnetic moment, $g-2$,
using a simple mass spectrum with gaugino mass unification.
In addition to the Higgsino-wino propagator,
there is also a contribution from the neutralino diagram with bino-bino propagator ($\tilde B$-$\tilde B$).
For the Higgsino-wino (and Higgsino-bino) contribution, the Higgsino vertex contains the Yukawa coupling
of muon/electron,
while for the bino-bino contribution, the left-right smuon/selectron mixing contains the muon/electron mass.
As a result, 
if there is no flavor violation,
the amplitude is proportional to the muon/electron mass in any diagrams
and the lepton mass scaling in Eq.(\ref{LMS}) is observed for $g-2$. 

\begin{figure}[htbp]
\center
\includegraphics[width=10cm]{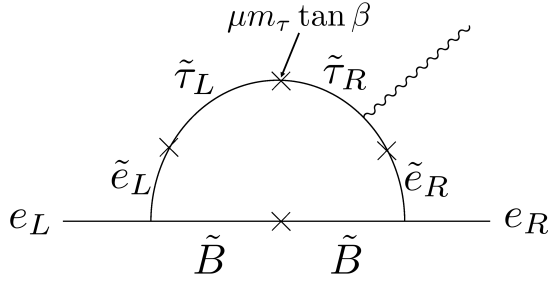}
\caption{
The neutralino (bino-bino) diagram to generate electron $g-2$.
The 1-3 flavor violation in left- and right-handed slepton masses
can break the lepton mass scaling of $g-2$ in Eq.(\ref{LMS}).
}
\label{fig1}
\end{figure}

We introduce the 1-3 flavor violation in the slepton mass matrices to
break the lepton mass scaling.
The lepton flavor violating decay $\mu \to e\gamma$ process has a strict experimental bound,
and we do not introduce any 1-2 flavor violation.
The coexistence of the 1-3 and 2-3 flavor violations can induce $\mu \to e\gamma$,
and thus we introduce only 1-3 flavor violation.
Under this assumption, the muon $g-2$ is generated from the diagonal elements in the slepton masses 
(without any flavor change) by
chargino and neutralino
 loop diagrams.
On the other hand,
if there are both left- and right-handed 1-3 flavor violation,
the neutralino loop diagram for electron $g-2$ can contain the $\tau$ mass instead of the electron mass 
as shown in Fig.\ref{fig1},
and thus the lepton mass scaling can be violated.

Using the conventional mass insertion parametrization,
\begin{equation}
\delta_{LL}^{13} = \frac{(m^2_{\tilde \ell}){}_{LL}^{13}}{m_{\tilde \ell}^2},
\quad
\delta_{RR}^{13} = \frac{(m^2_{\tilde \ell}){}_{RR}^{13}}{m_{\tilde \ell}^2},
\quad
\delta_{LR}^{13} = \frac{(m^2_{\tilde \ell}){}_{LR}^{13}}{m_{\tilde \ell}^2},
\end{equation}
where $m^2_{\tilde \ell}$ is an average slepton mass,
one can express the contribution to the electron $g-2$ via the $\tilde B$-$\tilde B$ diagram
as
\begin{equation}
(\delta_{LL}^{13} \delta_{LR}^{31} + \delta_{LR}^{13} \delta_{RR}^{31}) +\delta_{LL}^{13} \delta_{LR}^{33} \delta_{RR}^{31},
\end{equation}
where $\delta_{LR}^{33} = (A_\tau - \mu \tan\beta) m_\tau/m_{\tilde \ell}^2$.
We note that $\delta_{RL}^{31} = (\delta_{LR}^{13})^*$, $\delta_{LL}^{31} = (\delta_{LL}^{13})^*$, etc. are satisfied
due to the hermiticity of the slepton mass matrix\footnote{
Since the experimental bound of the electron electric dipole moment (eEDM) is quite severe,
we assume that all the elements in the slepton mass matrix are real (as we will comment later).}.
One can find that
the lepton mass scaling can be violated
with and without $\delta_{LR}^{13,31}$ flavor violation 
(induced by the scalar trilinear coupling $A$ which is not proportional to the charged lepton Yukawa matrix).
By choosing the signs of the off-diagonal elements, one can obtain the opposite sign of $\Delta a_e$ compared to $\Delta a_\mu$.
The magnitude of the electron $g-2$ discrepancy, $\Delta a_e$, is about 10 times larger compared to the 
one which is expected from $\Delta a_e \simeq m_e^2/m_\mu^2 \Delta a_\mu$ by the lepton mass scaling.
If the bino-bino contribution saturates the muon $g-2$, one  estimates that
$|\delta_{LL}^{13} \delta_{RR}^{31}| \sim 10 m_e/m_\tau \sim 0.05^2$
can realize the magnitude of electron $g-2$.
Usually, the bino-bino contribution is subdominant to muon $g-2$, and we need larger flavor violation to obtain the 
central value of electron $g-2$, but one can expect that $|\delta_{LL,RR}^{13}| \sim O(0.1)$ can reproduce 
the electron $g-2$.

The  Br($\tau \to e\gamma$) \cite{Aubert:2009ag} provides a constraint to achieve the central value of the electron $g-2$
by the flavor violation:
\begin{equation}
{\rm Br} (\tau \to e\gamma) < 3.3 \times 10^{-8}.
\end{equation}
In other words,
 $\tau \to e\gamma$ may be observed soon (but $\tau \to \mu\gamma$ will not)
if this realization of the electron $g-2$ is true.

The $\tau \to e\gamma$ amplitudes can be expressed by the mass insertion method as follows:
\begin{enumerate}
\item
$\tau_L \to e_R \gamma$
\begin{equation}
(\delta_{LR}^{31} + \delta_{LR}^{33} \delta_{RR}^{31})A_{\tilde B - \tilde B} + \delta_{RR}^{31} A^L_{\tilde H - \tilde B} .
\end{equation}

\item
$\tau_R \to e_L \gamma$
\begin{equation}
(\delta_{RL}^{31} +\delta_{RL}^{33} \delta_{LL}^{31})A_{\tilde B - \tilde B} + \delta_{LL}^{31} A^R_{\tilde H - \tilde W(\tilde B)} .\end{equation}

\end{enumerate}
If the $LR$ flavor violation is turned on,
the $\delta_{LR}^{31,13}$ can tune the amplitudes to satisfy the experimental bound of $\tau \to e\gamma$.
When the muon $g-2$ anomaly is satisfied (for gaugino and Higgsino masses: $M_1,M_2,\mu>0$),
one can obtain negative $\Delta a_e$
by choosing the signs of the off-diagonal elements as
\begin{equation}
\delta_{LL}^{13} : \pm, \quad 
\delta_{LR}^{13} : \mp, \quad
\delta_{RR}^{31} : \pm, \quad 
\delta_{LR}^{31} : \mp. \quad
\end{equation}

Even without any $LR$ flavor violation,
negative $\Delta a_e$ can be generated by $\delta_{LL}^{13} \delta_{RR}^{31} \neq 0$.
%
In this case, however, 
the contributions to the $\tau \to e\gamma$ amplitude have to be cancelled 
between bino-bino diagram and Higgsino-wino(bino) diagrams,
and thus the SUSY mass spectrum is constrained.
The bino-bino contribution to the amplitudes for both $\tau_L \to e_R \gamma$
and $\tau_R \to e_L \gamma$ behaves as
\begin{equation}
A_{\tilde B-\tilde B} \propto \alpha_Y \frac{M_1 \mu}{m_{\tilde \ell_L}^2 m_{\tilde \ell_R}^2} f_N (m_{\tilde \ell_L},m_{\tilde \ell_R},M_1),
\label{Bino-Bino}
\end{equation}
where $f_N$ stands for a loop correction for neutralino diagram.
The Higgsino-bino contribution to $\tau_L \to e_R \gamma$ amplitudes 
behaves as
\begin{equation}
A_{\tilde H-\tilde B}^L \propto -\frac{\alpha_Y}{M_1 \mu} f_N (M_1,\mu,m_{\tilde \ell_R}),
\label{Higgsino-Bino}
\end{equation}
and the behavior of the Higgsino-wino(bino) contribution to $\tau_R \to e_R \gamma$ amplitudes can be roughly
written as
\begin{equation}
A_{\tilde H-\tilde W(\tilde B)}^R \propto 
\frac{\alpha_2}{M_2 \mu} f_C (M_2,\mu,m_{\tilde \nu})
- \frac{\alpha_2}{2M_2 \mu} f_N (M_2,\mu,m_{\tilde \ell_L})
+\frac{\alpha_Y}{2M_1 \mu} f_N (M_1,\mu,m_{\tilde \ell_L}),
\end{equation}
where $f_C$ stands for a loop function for chargino diagram.
We find that $M_1/M_2<0$ is needed (where $M_1$ and $M_2$ are bino and wino masses)
and sleptons need to be enough light
to satisfy the experimental bound of $\tau \to e\gamma$ and to obtain the large magnitudes of $\Delta a_e$.
The opposite signs of $\Delta a_e$
compared to $\Delta a_\mu$ can be obtained by
$\delta_{LL}^{13} \delta_{RR}^{31} >0$.




\section{Numerical works}

In this section, we show our numerical calculations of  $g-2$ of muon and electron.
In the previous section, in order to illustrate the qualitative feature, we have used the mass insertion
approximation, but here, we calculate the $\bar\ell_i \sigma_{\mu\nu} \ell_j F^{\mu\nu}$ operator
 without using the mass insertion approximation.

\begin{figure}[tbp]
\center
\includegraphics[width=8cm]{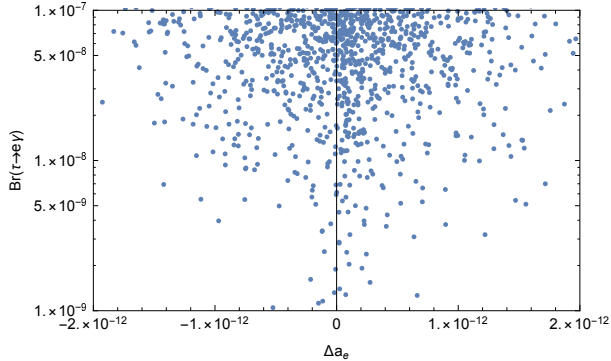}
\caption{The plot with $\delta_{LR}$ mass insertion.
$\delta_{LL,RR}^{13}$ and $\delta_{LR}^{13,31}$ are generated randomly.
The mass parameters are fixed
as $m_{L}=m_R =450$ GeV,  $M_1 = 310$ GeV, $M_2 = 350$ GeV, $\mu = 450$ GeV for $\tan\beta =40$
to realize the central value of muon $g-2$.}
\label{fig2}
\end{figure}

First, we show the case 
where $LR$ flavor violation is turned on.
As we have explained, the $\tau\to e\gamma$ amplitudes can be
canceled by choosing the $LR$ off-diagonal elements in this case.
As long as the  muon $g-2$ anomaly is satisfied, the mass spectrum does not get further constrained in order
to obtain the electron $g-2$ anomaly,
since the electron $g-2$ can be adjusted by choosing the  1-3 off-diagonal elements of the slepton mass matrices.
In this case, therefore, if the mass spectrum, which can reproduce the muon $g-2$, can satisfy the collider bounds,
the electron $g-2$ can be also reproduced without contradicting the experimental bounds in principle.
Since the main contribution to the electron $g-2$ emerges from the bino-bino diagram,
a heavier Higgsino mass $\mu$ and lighter sleptons and bino can reproduce the electron $g-2$ 
by smaller flavor violation.
Fig.\ref{fig2} shows the plots created by randomly generated $\delta_{LR}^{13,31}$ and $\delta_{LL,RR}^{13}$
for a fixed mass spectrum which can satisfy the muon $g-2$,
and one finds that the central value of the electron $g-2$ can be obtained
satisfying the experimental bound of $\tau\to e\gamma$.

Since we need to satisfy the $g-2$ of muon using the sparticles allowed by the LHC constraint, let us  discuss the LHC constraints on the non-colored sparticle masses~\cite{cmsstop,atlasstop} and the parameter space which is still allowed. We assume that the lightest neutralino is the lightest SUSY particle.  
\begin{enumerate}
\item
We first discuss  the slepton masses. 
The selectron and smuon masses do not have any constraint if the mass difference between the lightest neutralino and the selectron mass is $\leq 60$ GeV. Also, if the lightest neutralino mass is above 300 GeV then there is no constraint on the selectron and smuon masses. The stau masses do not have any constraint from the LHC yet.
\item
It is also interesting to note that if the lightest neutralino mass is above 300 GeV then there is no constraint on the next-to-lightest neutralino and chargino masses provided  the selectron and smuon masses are heavier than these particle masses. 
\item
If the lightest and next-to-lightest neutralinos and chargino are primary Higgsino then the maximum constraint on this mass scale  is 150-200 GeV. 
\item
In addition to the above straightforward scenarios, many other scenarios could be available by investigating the branching ratios (BR) of the heavy neutralinos and charginos into various final states, e.g., $W,\, Z,\, h,\, \tau,\,e, \,\mu,\, \nu$ plus $\tilde\chi^0_1$. The constraints shown by CMS are ATLAS mostly use  BR=1 for each of these final states.

\end{enumerate}


\begin{table}
\center
\begin{tabular}{|c|c|c|c|c|}
\hline 
 & Scenario 1  & Scenario 2  & Scenario 3 & Scenario 4 \\ \hline
 $M_1$ & 310 & 420 & 220 & 150 \\ 
 $M_2$ & 350 & 260 & 800 &300 \\ 
 $\mu$ & 450 & 250 & 230 & 620 \\ 
 $m_L=m_R$ & 450 & 530 & 300 & 540 \\ 
 $\tan\beta$ & 40 & 30 & 40 & 45 \\ 
\hline \hline
 $m_{\tilde \chi_1^0}$ & 301 & 197& 192& 149 \\ 
 $m_{\tilde \chi_2^0}$ & 332& 257& 235& 293\\ 
 $m_{\tilde \chi_3^0}$ & 455& 312& 254& 625\\ 
 $m_{\tilde \chi_4^0}$ & 482& 428& 809& 632\\  \hline
 $m_{\tilde \chi_1^+}$ & 327& 202& 227& 293\\ 
 $m_{\tilde \chi_2^+}$ & 480& 320& 809& 634\\ \hline
 $m_{\tilde e_{1,2}}$ & 429, 461& 510, 543& 288, 312& 517, 548\\ 
 $m_{\tilde \mu_{1,2}}$ & 450, 454 & 531, 533 & 302, 305 & 539, 544 \\ 
 $m_{\tilde \tau_{1,2}}$ & 406, 507 & 493, 578 & 266, 343 & 487, 608\\ 
 $m_{\tilde \nu}$ & 417, 445, 472 & 499, 526, 552  & 271, 293, 314 & 506, 536, 564 \\ 
 \hline
 $\Delta a_\mu \times 10^9$ & 2.8 & 2.7 & 2.9& 2.4 \\ \hline
\end{tabular}
\caption{The mass spectrum for gaugino masses, $M_1$, $M_2$,
Higgsino mass $\mu$, SUSY breaking left- and right-handed slepton masses, $m_L$, $m_R$,
and ratio of the Higgs vacuum expectation values, $\tan\beta = v_u/v_d$,
 in the scenarios given in the text.
The electron $g-2$ is adjusted to the central values by assuming $\delta_{LL}^{13} = \delta_{RR}^{13}$,
and $\tau\to e\gamma$ is cancelled by the freedom of $\delta_{LR}^{13,31}$.
The selectron and sneutrino masses are split due to these off-diagonal elements.
We note that the mass eigenstates of the selectron (stau) contain stau (selectron) of the current basis at $O(10) \%$.
We choose $A_\tau = 0$.}
\label{Table1}
\end{table}

Following these prescriptions, we are showing 4 points which are not ruled out by the LHC data
in Table \ref{Table1}. The lightest neutralino are chosen to be  Higgsino, wino-Higgsino, bino-Higgsino or bino types to show different possibilities.


\begin{itemize}
\item  Scenario 1:

All  the heavier neutralinos or charginos dominantly decay via $W$, $Z$, $\tau$$+\tilde\chi^0_1$. LHC constraints are satisfied since the lightest neutralino mass, $m_{\tilde{\chi}_{1}^0}$, is 300 GeV. There exists no constraint on the selectron, smuon masses since $m_{\tilde{\chi}_{1}^0}$ is 300 GeV and  all the LHC constraints are satisfied.

\item Scenario 2: 


The  lightest  neutralino and the lightest  chargino are within 10 GeV and they are around 200 GeV. The lightest chargino and the neutralino masses are   required to be above 160 GeV  in such a degenerate case~\cite{Sirunyan:2018iwl, Aaboud:2017leg}. Two other neutralinos and the heaviest charginos are within 120 GeV of the lightest neutralino and they decay dominantly via $W$, $Z,\, h$+$\tilde\chi^0_1$. In such final states the mass difference between the heavier neutralino/chargino and the lightest neutralino needs to be at least 200 GeV for $m_{\tilde\chi^0_1}\sim 200$ GeV in order to have any constraint from the LHC.  The lightest neutralino is wino-Higgsino type.  The  heaviest neutralino is more than 95\% bino, which would make it hard to be produced at the LHC. There is no constraint on the selectron and smuon  masses above 500 GeV for $m_{\tilde\chi^0_1}\sim 200$ GeV.  All the LHC constraints are satisfied for this scenario. 

\item Scenario 3:

Three lighter neutralinos and the lightest chargino are within 60 GeV and the lightest neutralino is bino-Higgsino type. These heavier particles decay  into the lightest neutralino  via $W^\ast,\,Z^\ast$ and there exists no constraint on these particle since for a lightest neutralino around 200 GeV, the mass difference is needed to be at least 200 GeV for $W,\ Z$ final sates to have constraints from the LHC.   The heaviest neutralino and the chargino are wino-type with mass around 800 GeV. The heaviest neutralino decays into $\nu_L\tilde{\nu}_L$  (36\%), $\tau\tilde\tau$ (12\%), $l_L\tilde{l}_L$ (24\%) where $l=e,\mu$ and $\tilde\nu_L$ decays mostly into $W^\ast+\tilde\chi^0_1+\tau$ and $\tilde\chi^0_1+\nu$ and the heavy chargino decays into $l_L\nu_L+\tilde\chi^0_1$ (35\%), $\tau\nu+\tilde\chi^0_1$ (10\%) and others. One can find that the the $lll\nu$ final state cross-section is around 0.02fb which is quite small compared to the cross-section ($\sim$ 1 fb) that can be constrained for this final state for a 800 GeV $\tilde\chi^0_4,\,\tilde\chi^\pm_2$ with $m_{\tilde{l}}=0.05 m_{\tilde\chi^0_1}+0.95 m_{\tilde\chi^\pm_2}$ and a 200 GeV $\chi^0_2$~\cite{Sirunyan:2017lae}. The  $\tilde\chi^0_4,\,\tilde\chi^\pm_2$ masses are too large to be constrained by any other final state.  The selectron and smuon masses do not have constraint for masses up to 330 GeV for $m_{\tilde\chi^0_1}\sim 200$ GeV. The scenario 3, therefore, cannot be constrained by the LHC results so far.   
 
\item  Scenario 4:

The selectron and smuon masses are  heavier than the two lighter neutralino and the lightest  chargino masses. The final states of the $\tilde{\chi}_{ 2}^0/\tilde{\chi}_{1}^\pm$ will be dominated by $W$, $Z$ and there exists no constraint on these masses up to 300 GeV for $m_{\tilde\chi^0_1}\sim 150$ GeV. The heavier neutralinos and charginos are primarily Higgsinos in such a scenario and they decay via $W, \,Z,\, h$. Their masses are heavier than 600 GeV which leads to  no constraint from the LHC. There is no constraint on the selectron and smuon  masses above 500 GeV for $m_{\tilde\chi^0_1}\sim 200$ GeV. The LHC constraints do not apply to this scenario

\end{itemize}

\begin{figure}[tbp]
\center
\includegraphics[width=8cm]{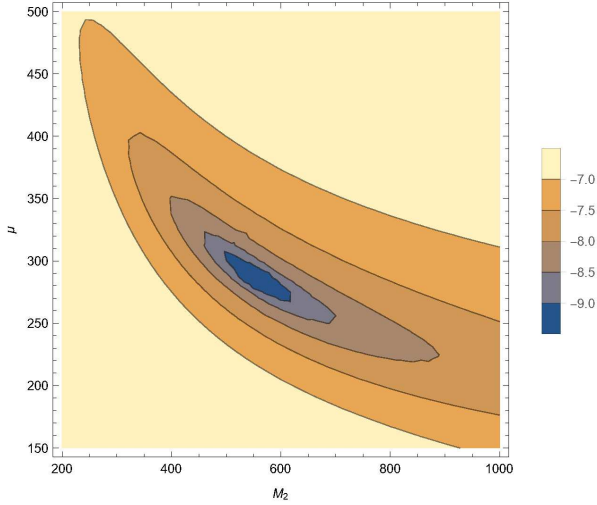}
\caption{
Contour plot of $\log_{10} {\rm Br}(\tau\to e\gamma)$ on the $M_2$-$\mu$ plane.
We fix the parameters as
$M_1 = -220$ GeV, $m_L = m_R= 300$ GeV,
$\tan\beta = 40$, $A_\tau = -3$ TeV, and $\delta_{LL,RR}^{13} =0.1$.
}
\label{fig3}
\end{figure}

Let us now present the case of $LR$ mixing is 0. In the case where $LR$ flavor violation is absent,
the mass spectrum is constrained to satisfy the $\tau\to e\gamma$ bound.
Fig.3 shows the contour plot of ${\rm Br}(\tau\to e\gamma)$
fixing bino mass $M_1$ and slepton masses, $m_L$ and $m_R$.
As we have mentioned, $M_1/M_2$ has to be negative 
to cancel the amplitudes of $\tau\to e\gamma$.
Since the chargino contribution dominates the muon $g-2$ in the
region where $\tau\to e\gamma$ is suppressed,
the smaller $M_2$ and $\mu$ can generate larger muon $g-2$.
On the other hand, the magnitude of the electron $g-2$ is 
dominated by the bino-bino diagram, and it 
can be larger for larger $\mu$
and larger flavor violation.
When we choose the flavor violation to reach the central value of 
 electron $g-2$ and select the mass spectrum to cancel $\tau\to e\gamma$
 amplitudes, we find that it is not easy to reach the central value of muon $g-2$
 (as far as the smuon mass is the same as the average of selectron and stau masses),
 but one can find a solution to achieve the $1\sigma$ range of the muon $g-2$. 
 From the cancellation of $\tau_L \to e_R \gamma$ amplitude between Eqs.(\ref{Bino-Bino}) and (\ref{Higgsino-Bino}),
 we need $M_1 \mu \sim m_{\tilde \ell}^2$ naively.
 If $A$-term coefficient is large, the Higgsino mass can be lowered 
 to cancel the $\tau_L \to e_R \gamma$ amplitude and enlarge muon $g-2$. 
 (A large $A$-term coefficient can lead the existence of charge breaking global minimum,
  but it can be avoided if the CP odd Higgs mass is adequately heavy (roughly more than $|A_\tau|/\sqrt3$) in this case).
 In the example given in Fig.\ref{fig3}, 
 we obtain $\Delta a_\mu = 2.3 \times 10^{-9}$ at $M_2 = 800$ GeV and $\mu = 230$ GeV,
 and the electron $g-2$ can reach the central value, $\Delta a_e = -8.7 \times 10^{-13}$,
 satisfying the $\tau \to e\gamma$ bound.

We show one typical  benchmark point for the chargino, neutralino, and slepton spectrum
which can satisfy the $\tau\to e\gamma$ bound and reach the $1\sigma$ range of 
$g-2$:
\begin{equation}
M_1 = -220 \ {\rm GeV}, \ M_2 = 800\ {\rm GeV}, \ \mu = 230\ {\rm GeV}, \ m_L= m_R = 300\ {\rm GeV},
\ \tan\beta=40,
\end{equation}
 and $A_\tau = -3$ TeV, in the convention that the $LR$ component of slepton mass matrix
is $(A_\tau - \mu \tan\beta) m_{\tau}$. 
The mass spectrum is
\begin{equation}
m_{\tilde\chi_{1,2,3,4}^0} = 197, \ 226, \ 258, \ 809 \ {\rm GeV}, \qquad
m_{\tilde\chi^+_{1,2}} = 227, \ 809 \ {\rm GeV},
\end{equation}
\begin{equation}
m_{\tilde e_{1,2}} = 289, \ 317  \ {\rm GeV}, \qquad
m_{\tilde \mu_{1,2}} = 301, \ 305  \ {\rm GeV},  \qquad
m_{\tilde \tau_{1,2}} = 250, \ 349  \ {\rm GeV},
\end{equation}
\begin{equation}
m_{\tilde \nu} = 265, \ 293, \ 319 \ {\rm GeV}.
\end{equation}
This scenario is similar to scenario 3 described above and therefore is not constrained by any LHC data.

We note that the electron mass is modified by finite loop correction
at $O(10)$\% when the central value of the electron $g-2$ is reproduced.

Before concluding this section, we note on the bounds of electron EDM \cite{Andreev:2018ayy}:
\begin{equation}
|d_e| < 1.1 \times 10^{-29} \, e \cdot {\rm cm}.
\end{equation}
Both $g-2$ and EDM of electron is generated by the $\bar e_L \sigma_{\mu\nu} F^{\mu\nu} e_R$ operator,
and thus, if the SUSY contribution saturates the deviation from the SM prediction of $g-2$,
we obtain
\begin{equation}
d_e^{\rm SUSY} = \frac{e}{2m_e} a_e^{\rm SUSY} \tan\phi
\simeq  2 \times 10^{-11} \Delta a_e \tan\phi \ e\cdot {\rm cm},
\end{equation}
where $\phi$ is a phase of the amplitude.
We usually suppose that the gaugino, Higgsino mass and the (diagonal elements of) scalar 
trilinear coupling matrix are real to satisfy the electron and neutron EDM bounds.
The phases of the off-diagonal elements also need to be almost real,
and the bound of the phases is
\begin{equation}
|{\rm arg}\,\delta_{LL,RR,LR,RL}^{13} |< O(10^{-6}).
\end{equation}
If the phases are aligned as
\begin{equation}
{\rm arg}\,\delta_{LL}^{13} = {\rm arg}\,\delta_{RR}^{13} = {\rm arg}\,\delta_{LR}^{13} =
{\rm arg}\,\delta_{RL}^{13},
\end{equation}
the phases are unphysical since they can be removed by field redefinition, and the electron EDM
vanishes.

\section{Conclusion}

In conclusion, we have investigated the recently reported more than $2 \sigma$ electron $g-2$ anomaly. The discrepancy between the SM prediction and the experimental measurement for the electron case has opposite sign compared to the muon $g-2$ case where the anomaly is more than $3 \sigma$. Further, the ratio of the measured  muon and electron anomalies is about 10 times less than that predicted by the lepton mass scaling $m_\mu^2/m_e^2$. One  requires flavor violation in the leptonic sector to induce such a breakdown of the scaling.

In this work we showed that it is possible to explain electron and muon $g-2$ anomalies simultaneously in the MSSM using the parameter space which is allowed by the LHC data. The satisfaction of the $g-2$ anomalies require non-colored particles, such as selectrons, smuons, staus, neutralinos and charginos, and the LHC constraints on these particles allow sufficient parameter space for masses between 100 GeV - 1 TeV.  Further,   while satisfying electron $g-2$, we found that flavor violating $\tau\rightarrow e\gamma$ decay is induced which, however, can be lower than the current experimental constraints.  

\section*{Acknowledgments}


The work of B.D. is supported by DOE Grant de-sc0010813.
The work of Y.M. is supported by Scientific Grants 
by the Ministry of Education, Culture, Sports, Science and Technology of Japan 
(Nos. 16H00871, 16H02189, 17K05415 and 18H04590). We thank H. Baer,  T. Kamon, T. Li, K. Olive and C. Wagner for discussions.

\end{document}